\documentstyle[epsf,prd,aps,floats]{revtex}
\input{psfig.tex}
\newcommand{\be}{\begin{equation}}
\newcommand{\ee}{\end{equation}}
\newcommand{\bea}{\begin{eqnarray}}
\newcommand{\eea}{\end{eqnarray}}
\tighten
\begin{document}
\draft
\preprint{Imperial/TP/00-01/??
}
\twocolumn[\hsize\textwidth\columnwidth\hsize\csname
@twocolumnfalse\endcsname

\title{Nielsen--Olesen vortex in varying-$\alpha$ theories}
\author{Jo\~ao Magueijo, H\aa vard Sandvik and T.W.B. Kibble}
\address{Theoretical Physics, The Blackett Laboratory,
Imperial College, Prince Consort Road, London, SW7 2BW, U.K.}
\date{\today}
\maketitle
\begin{abstract}
We consider soliton solutions to Bekenstein's theory, for which the
fine structure constant $\alpha=e^2/(4\pi\hbar c)$ is allowed to vary
due to the presence of a dielectric field pervading the vacuum.
More specifically we investigate the effects of a varying $\alpha$
upon a complex scalar field with a $U(1)$ electromagnetic gauge symmetry
subject to spontaneous symmetry breaking.  We find vortex solutions
to this theory, similar to the Nielsen--Olesen vortex. Near the vortex
core the electric charge is typically much larger than far away from the
string, lending these strings a superconducting flavour. In general the
dielectric field coats the usual local string with a global string
envelope.
We discuss the cosmological implications of networks of such strings,
with particular emphasis on their ability to generate inhomogeneous
recombination scenarios.
We also consider the possibility of the dielectric being a charged free
field. Even though the vacuum of such a field is trivial, we find that
the dielectric arranges itself in the shape of a local string, with a
quantized magnetic flux at the core --- presumably borrowing these
topological features from the underlying Nielsen--Olesen vortex.
\end{abstract}

\date{\today}

]

\renewcommand{\thefootnote}{\arabic{footnote}}
\setcounter{footnote}{0}

\section{Introduction}
The possibility that the constants of Nature may not be
constant at all has long been entertained by physicists
(see for instance the groundbreaking work of Dirac
\cite{dirac}). Historically the gravitational
constant $G$ has been the main victim of attempts to build
varying constant theories. An example of a varying-$G$ theory
is Brans--Dicke theory \cite{dicke}, but modern string theories all
display
this property (eg. \cite{gibbons}). Another often dethroned constant is
the
electron charge $e$, whose variability is envisaged by
varying-$\alpha$ theories (where $\alpha=e^2/(4\pi\hbar c)$
is the fine structure constant)
such as the one proposed by Bekenstein \cite{bek2}. A recent claim
for observational evidence for a time-varying $\alpha$
\cite{webb,va1} has
sparked great interest in this type of theories
\cite{landau,avel,bat}. More radically a
varying $\alpha$ may result from  varying speed
of light (VSL) theories, typically
(but not always \cite{vslcov}) entailing
breaking of Lorentz invariance \cite{mof,am}.
It appears that VSL theories may solve
the cosmological problems usually solved by inflation.

Should the observations of Webb et al be confirmed, no doubt
most of the physical constructions employed by physicists
will have to be reexamined. Bekenstein's theory is perhaps the
most conservative theory with which to interpret the new results,
in the sense that it does not require giving up any truly fundamental
principles, such as covariance and Lorentz invariance. Within
the framework of this theory the vacuum is pervaded by a dielectric
medium,
screening the electric charge. The properties of this dielectric
medium are
determined by the electromagnetic field itself, within the context
of a dynamical Lagrangian theory. Hence in the surroundings
of any object with an electromagnetic energy component, there
will be spatial variations in the electric charge.

The application of this theory to cosmology is clouded by the issue
of determining how much of the matter in the Universe will act as a
source for this dielectric medium \cite{bek2} (see however
\cite{prep}). Clearly one needs to understand
the microphysics underlying the cosmological fluid, in particular
the nature of the dark matter,  in order to set
up a consistent cosmological model. In this paper we therefore turn
to a more concrete and simpler situation. We consider a complex scalar
field with a $U(1)$ gauge electromagnetic symmetry spontaneously
broken, which couples
to a dielectric field in accordance with Bekenstein's prescription. We
then
consider topologically non-trivial solutions to this theory,
the counterpart of the Nielsen--Olesen vortex \cite{niel}. In the standard
theory such vortices have well localized concentrations of energy,
along a stable string-like core. Furthermore this core constitutes
a magnetic field flux tube. Hence the vortex acts as a source for
the dielectric vacuum proposed  in \cite{bek2}, leading to a varying
$\alpha$ in the vicinity of the string.

The value of $e$ in the string core is therefore much larger (smaller)
than the asymptotic value $e_0$ (larger or smaller depending on
parameter signs). If $e$ becomes much larger we obtain a situation
vaguely similar to the superconducting strings of Witten \cite{witten}.
Indeed in some sense, infinite charge could amount to superconductivity.
In Sec.~\ref{model} we set up the formalism, and study the asymptotics
of our solution, and in Sec.~\ref{numsol} present the full numerical
solution. A qualitative discussion of the implications of cosmological
networks of such strings is presented in Sec.~\ref{cosmo}. Such strings
would combine local and global string elements in their evolution and
energy loss mechanisms, as well as in their gravitational interaction
with the surrounding matter. More distinctly they would generate
inhomogeneities in the value of $e$, leading, among other things,
to inhomogeneous recombination scenarios.

Another interesting connection, spelled out
in Sec.~\ref{fast},  is the similarity between our solutions
and fast-tracks, a construction found in VSL theories \cite{vslcov}.
Such solitonic solutions to VSL allow for fast travel without a
time-dilation effect. We discuss how the situation is distinctly
different in the case of these strings --- they still allow for
fast travel in some sense, however they would induce a time-dilation
effect of their own which has nothing to do with the special relativity
effect.

The  solution derived in Sec.~\ref{model}
is but the simplest of many similar constructions
involving solitonic solutions coupled to varying charge theories. In all
of
these a gauge field undergoing spontaneous symmetry breaking
supplies a solitonic solution which acts as a source for a dielectric
field. As a result a dielectric coating is superposed on the soliton,
forcing the gauge coupling (or charge) to vary in the soliton
core or near its vicinity.  In Sec.~\ref{diegauge}
we discuss the possibility
of gauging the dielectric field itself. In a concluding
discussion, in Sec.~\ref{conc} we also consider the possibility
of non-Abelian gauge groups, and similar constructions with the morphology
of monopoles and textures.

\section{The model}\label{model}
We first describe Bekenstein's theory, in the context of a charged
complex scalar field undergoing spontaneous symmetry breaking.
Let $\phi$ be a complex scalar field with a gauged $U(1)$
symmetry, and $A_\mu$ be the gauge field. Let the electric
charge $e$ be a variable, with $\epsilon=e/e_0$ where $e_0$ is
some fixed electric charge. Under a gauge transformation
$\delta\phi=i\Lambda\phi$, where $\Lambda$ is a scalar function,
one should impose $\delta A_\mu
=-(\partial_\mu\Lambda)/e$, so that the derivative
$D_\mu=\partial_\mu +ieA_\mu$ transforms covariantly. The
gauge invariant electromagnetic field tensor is then
\be
F_{\mu\nu}={1\over \epsilon}(\partial_\mu(\epsilon A_\nu)
- \partial_\nu(\epsilon A_\mu) )
\ee
and the Lagrangian is:
\be
{\cal L}=-(D^\mu\phi)^\star D_\mu\phi-V(\phi)
-{1\over 4}F_{\mu\nu}F^{\mu\nu} - {\omega\over 2 \epsilon^2}
\partial_\mu\epsilon\partial^\mu\epsilon
\ee
(we are using a metric with signature $-+++$).
The first three terms constitute the matter Lagrangian, while
the last term governs the dynamics of $\epsilon$.
We adopt the proverbial Mexican hat potential:
\be
V(\phi)=m^2|\phi|^2+\lambda|\phi|^4,
\ee
with $\lambda$ and  $m^2<0$ fixed parameters.
The vacuum is then the circle
$|\phi|=\phi_0={\sqrt {-m^2/(2\lambda)}}$.

We first introduce a transformation which simplifies
Bekenstein's theory enormously. We note that by defining
an auxiliary gauge potential $a_\mu=\epsilon A_\mu$ and
field tensor:
\be
f_{\mu\nu}=\epsilon F_{\mu\nu} =\partial_\mu a_\nu
-\partial_\nu a_\mu
\ee
the Lagrangian becomes
\be \label{l2}
{\cal L}=-(D^\mu\phi)^\star D_\mu\phi-V(\phi)
-{1\over 4}{f_{\mu\nu}f^{\mu\nu}\over \epsilon^2}
 - {\omega\over 2 \epsilon^2}
\partial_\mu\epsilon\partial^\mu\epsilon,
\ee
in which $D_\mu=\partial_\mu +ie_0 a_\mu$. Hence we have eliminated
the dependence on $\epsilon$ in the matter Lagrangian
apart from dividing the $f^2$ term
by $\epsilon^2$. This greatly simplifies the variational
problem regardless of which variables we decide to label as physical
(which is essentially a matter of convention). Indeed zero variation
with respect to $\{\phi,A_\mu,\epsilon\}$ is equivalent to zero
variation with respect to $\{\phi,a_\mu,\epsilon\}$.
We have also exposed an interesting similarity between this theory
and Brans--Dicke changing-$G$ theory. In the latter one multiplies
the Ricci scalar (essentially a $f^2$ term) by a scalar field, which
also does not appear explicitly elsewhere (other than in its own
kinetic terms or potential).

Variation of (\ref{l2}) with respect to $\phi$ produces the equation:
\be
D_\mu D^\mu\phi={\partial V\over \partial \phi^\star},
\ee
in which we may use $D_\mu=\partial_\mu +ie_0 a_\mu$
or $D_\mu=\partial_\mu +ie A_\mu$. Variation with respect
to $a_\mu$ now produces straightforwardly:
\be
\partial_\mu{f^{\mu\nu}\over \epsilon^2}=
\partial_\mu{F^{\mu\nu}\over \epsilon}=j^\nu=ie_0[\phi^\star
D^\nu\phi - \phi (D^\nu\phi)^\star]
\ee
and finally variation with respect to $\epsilon$
leads to:
\be
\Box\ln\epsilon =-{1\over 2\omega}{f^2\over \epsilon^2}
=-{1\over 2\omega}F^2.
\ee
These equations, in the $\{A_\mu,F_{\mu\nu}\}$ representation,
are nothing but Bekenstein's equations. However
the transformation we have used has simplified the derivation
greatly, and it will also simplify the search for solutions
in what follows.

We now seek solutions similar to the Nielsen--Olesen vortex in this
theory. We therefore introduce the ansatz $\phi=\chi(r) e^{in\theta}$
and $a_\theta=a(r)$ with all other components for $a_\mu$ set to
zero. We define a magnetic field out of the $f_{\mu\nu}$ tensor,
so that ${\bf b}=\nabla\times {\bf a}$. Hence the magnetic field
is aligned with the $z$ direction and has value:
\be\label{eq0}
b=b_z={1\over r}{d\over dr}[ra].
\ee
The dynamical equations are:
\be\label{eq1}
{1\over r}{d\over dr}{\left(r{d\chi\over dr}\right)}
-{\left[ {\left({n\over r} - e_0a\right)}^2 + m^2 + 2\lambda\chi^2
\right]}
\chi=0,
\ee
which is unmodified, and
\be\label{eq2}
{d\over dr}{\left( {1\over r\epsilon^2}{d\over dr}
(ra)\right)} + 2e_0{\left( {n\over r}- e_0 a\right)}\chi^2=0
\ee
and
\be\label{eq3}
{1\over r}{d\over dr}{\left(r{d\ln \epsilon \over dr}\right)}
=-{1\over 2\omega}{b^2\over \epsilon^2}.
\ee

To investigate the asymptotic behaviour far from the core, first
recall that the scalar field takes on the constant value $\chi =
\sqrt{-\frac{m^2}{2\lambda}}$ for $r\rightarrow \infty$.  From
Eq.~(\ref{eq1}) we then see that
\be a={n\over r e_0},
\label{asymp2}
\ee
which also agrees with Eq.~{\ref{eq2}.  From
this we deduce that the flux of $b$ is quantized:
\be
\int{\bf
b}\cdot {\bf dS}= \oint {\bf a} \cdot {\bf dl}={2\pi n\over e_0}
\ee
but not the flux of $B=b/\epsilon$ in which $\epsilon$ takes the
role of a magnetic permeability.  We also find the asymptotic
solution for $\epsilon$:
\be \epsilon={\left( r\over
r_0\right)}^{-I\over \omega\pi}, \label{asymp3}
\ee
where $I$ is the integral of ${\bf B}^2$ over a string cross
section. Hence $\epsilon$ can only either go to zero (if
$\omega>0$) or infinity (if $\omega<0$) far away from the string
--- corresponding to a logarithmic divergence in $\psi$; not
surprising since (\ref{eq3}) has sources at spatial infinity (at
$z=\pm\infty$). Furthermore the energy in the $\epsilon$ field
diverges away from the string. These features are well known
properties of global strings. Indeed we have a local or gauged
string (in the field $\phi$) superposed on a global string (in the
field $\epsilon$).

\begin{figure}
\begin{center}
\psfig{file=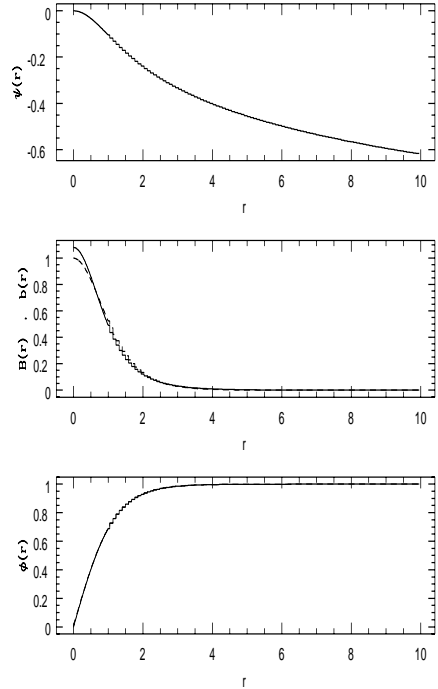,width=6cm,height=10cm} \caption{Numerical
integration of the Nielsen--Olesen vortex.  The first graph shows
the solution of $\psi= ln \epsilon$ as a function of $r$, in
Bekenstein's theory.  The middle graph shows a plot of $B$ with a
constant electric charge (dashed line) compared to a plot of $b$
in Bekenstein's theory.  The third graph compares the solution of
the $\phi$ field in the two theories; the solutions are
indistinguishable.}
\label{nielsen2}
\end{center}
\end{figure}

In a cosmological
setting these seemingly pathological divergences are naturally
removed by the scale of curvature of the strings. Then the difference
between the asymptotic and core values of the electric charge is
roughly of the order of $(r_c/r_0)^{-I/ (\omega\pi)}$, where
$r_c$ and $r_0$ are the curvature and core radii of the string
respectively.

If we require that $\psi$ has a positive definite energy, then
$\omega>0$ in which case the charge at the core should be much higher
than its asymptotic value. It is in this case that we can claim
a similarity between our construction and superconducting strings
\cite{witten}. In some loose sense a diverging electric charge
should be equivalent to superconductivity. Indeed applying an electric
field upon a conductor in the interior of this string subjects the
charge carriers to a force proportional to $e$. Hence the electric
force applied to them is much larger than normal. If the resistance
to which they are subject does not change,  we can then ignore it
--- and it is in this sense that these strings could maintain
persistent currents and therefore be labelled superconducting.
Note that this is just a loose analogy: effects such as the expulsion
of magnetic fields from the interior of the string are not present in
this case.

Our solution also has vague similarities to the dilatonic string
of \cite{dilaton}, for which the string mass per unit length is
a function of a scalar field.

\section{Numerical Solutions of the Model}
\label{numsol}
Equations (\ref{eq1})--(\ref{eq3}), together with the asymptotic values
at $r=0$ and $r=r_c$, constitute a boundary value
problem.  For the sake of numerically solving this problem, it is
convenient to make the following change of variables so as
to avoid singularities:
\bea
 a &\rightarrow &v = a r,\\
 \epsilon &\rightarrow & \psi = \ln \epsilon.
\eea
We also reduce the problem to first order by introducing the new
variables
\be
    \sigma = \frac{d \chi}{dr}, \qquad \eta = \frac{d \psi}{dr}.
\ee
The new set of equations suitable for numerical implementation is
\bea
 \frac{d \chi}{dr}& = &\sigma,
\label{numeq1}\\
 \frac{d \sigma}{dr}& = & - \frac{\sigma}{r} +
\left(\left(\frac{n-e_0v}{r}\right)^2 + m^2 +2\lambda \chi^2\right)\chi,
\label{numeq2}\\
  \frac{dv}{dr}&  = &b r,\\
  \frac{db}{dr} & = &
2\eta b - \left(2e_0 \frac{n - e_0v}{r} \chi^2\right)e^{2\psi}, \\
\frac{d \psi}{dr}& = &\eta,\\
\frac{d \eta}{dr}& = & -{\eta \over r} - { 1\over 2 \omega}
b^2e^{-2\psi}. \label{numeq6}
\eea

We first check our code on the Nielsen--Olesen vortex with
constant $\epsilon$.  The results for the scalar field and the
magnetic field compare well with the original work\cite{niel}, and
are shown as
the dashed lines in Figure~\ref{nielsen2}.  We then allow the
$\epsilon$ to vary, and incorporate
Equations~(\ref{numeq1})--(\ref{numeq6}).  The results are consistent with
the asymptotic behaviour found in
Equations~(\ref{asymp2})--(\ref{asymp3}),
and are shown as the solid lines in Figure~\ref{nielsen2}.

\section{Qualitative discussion of a cosmological network}
\label{cosmo}
Varying-$\alpha$ strings, if formed at phase transitions, would
have a complex evolution. It is conceivable that the string core
would still be governed by the Nambu--Goto action. Also,
presumably these strings, when crossing, would still intercommute
(although this fact should be checked by numerical
simulations). However, in addition to intercommuting,
the dielectric field would act as a long-range force between the strings,
creating a double mechanism for string interaction. Hence we
should have something like a local string network, acting as a
source for a global string network, the two networks being
driven by  their usual interaction mechanisms plus a complex
interaction between the two.

Energy loss
processes would again combine local and global string
elements. The core string
should develop small scale structure, via intercommutation, thereby
emitting gravitational radiation. On the
other hand the dielectric field would now supply a channel for the string
to lose energy via the emission of scalar
radiation.  A combination of processes peculiar to local and global
strings
should therefore push these strings towards a scaling solution, but
clearly
we may expect such scaling solutions to be distinctly different
from the usual ones.

The interactions of these strings and the other matter in the universe
would also be rather peculiar. Gravitationally we would find a combination
of the effects of local strings (and their conical flat space) and
the more complex global string gravitational fields. However, predicting
the density fluctuations in this scenario as a simple superposition
of global and local fluctuations (that is the total spectrum as a
weighted average of the separate spectra) is clearly too gross an
approximation. The local and global string networks will be highly
correlated, and have a strong feedback effect upon each other.
Whatever the gravitational effects of these strings upon the
surrounding matter will be, they have to be determined by simulations
along the lines of \cite{pst,cont} specifically applied to
varying-$\alpha$ string networks.
Note that unlike conventional
super-conducting cosmic strings we would not expect the equation of state
of these strings to differ from standard ones.

However what would truly distinguish these strings, and their
possible cosmological effects, is the fact that the
electric charge varies in their vicinity. For a straight string the
electron charge variation away from the core would be given by:
\be
{\Delta e\over e}={\beta\over \omega\pi}G\mu\ln {r\over r_0},
\ee
in which for the $I$ defined after equation \ref{asymp3}, we used
$I=\beta G\mu$. Here $\mu$ is the string mass per unit length, and
$\beta$ is the fraction of this mass in the magnetic field flux
tube. These fluctuations are therefore of the order of the
gravitational potential induced by the strings.

Hence, in addition to acting
as gravitational seeds for structure formation, these strings would
affect any electromagnetic processes in their neighbourhood. A topical
example is recombination. In the vicinity of these strings
the hydrogen binding energy would suffer spatial variations, leading
to inhomogeneous recombination. The implications of a
homogeneous changing-$\alpha$ upon recombination and CMB anisotropy
were studied in \cite{avel,bat}. A network of changing-$\alpha$ strings
would provide additional effects.

\section{A comparison of varying-$\alpha$ strings and fast tracks}
\label{fast}
Although there is a parallel between the solutions found in this
paper and fast-tracks \cite{vslcov}, there is a crucial difference.
Fast-tracks are string-like solutions to some covariant VSL theories
such that the speed of light is much higher near the string core.
Hence observers may move along the string core much faster than the
asymptotic value of $c=c_\infty$. Moreover such ``super-$c_\infty$''
speeds
need not be relativistic, that is, they may still be much smaller
than the local value of $c$, so that such observers
would not be subject to
time-dilation effects. Fast-tracks are what space travel is begging
for: fast, ``superluminal'' travel, free from time dilation. It can be
shown that a change of units transforms fast-tracks into wormholes.

In the case of our strings the situation is rather different.
As $\alpha$ changes near the string core so will change the time rates
associated with all electromagnetic processes. In particular an atomic
clock,
ticking to a rate $\tau\propto 1/\alpha^2$, will tick differently.
Biological processes, being electromagnetic, also tick to this rate.
If the charge decreases towards the string core, then alpha is smaller,
and so the time scales $\tau$ of all electromagnetic processes increase.
Unfortunately this situation is realized in the case $\omega<0$, for
which the dielectric energy density is not positive definite. Nonetheless
let us consider further this case.

         Using coordinate time we know that ``super-$c_\infty$'' speeds
cannot be achieved near the string. However, since $e\rightarrow 0$,
it would then be possible to ``pickle'' observers moving along the string,
since $\tau\rightarrow \infty$. If we were to measure speeds along the
string in atomic clock units, then we could indeed measure
``super-$c_\infty$'': the point is that natural organisms would be
able to travel very large distances within their perceived time scales.

However the use of such strings for space travel would still lead
to twin paradox effects: clearly a round trip
would cause a large difference in ages between
sedentary and nomadic twins. Curiously enough such a time dilation
effect has nothing to do with relativistic speeds. It is simply given
by
\be
\Delta t=\Delta t_0 \oint {dt\over \alpha^2}.
\ee
In this respect the varying-$\alpha$ strings considered here are
distrinctly different from VSL fast-tracks.

\section{Gauged dielectric field}\label{diegauge}
We have noted that the dielectric coating surrounding our modified
Nielsen--Olesen vortex is like a global string superposed on the
usual gauged string (made up of charged scalar field
and a magnetic flux tube). Further symmetry would be enforced if the
dielectric itself were charged, that is if we replaced Bekenstein's real
scalar field $\psi=\ln \epsilon$, by a complex field $\psi$, such that
$\epsilon=e^{|\psi|}$. Upon gauging the $U(1)$ symmetry associated
with $\psi$ we therefore arrive at the Lagrangian:
\bea \label{l3}
{\cal L}&=&-(D^\mu\phi)^\star D_\mu\phi-V(\phi)
-{1\over 4}{f_{\mu\nu}f^{\mu\nu}\over \epsilon^2}\nonumber\\
&& - \omega[
({\tilde D}_\mu\psi)^\star({\tilde D}^\mu\psi)
+{1\over 4}G_{\mu\nu}G^{\mu\nu}],
\eea
in which ${\tilde D}_\mu=\partial_\mu + igB_\mu$, where
$g$ the charge of the dielectric field, $B^\mu$ is the photon
associated with the dielectric, and $G_{\mu \nu} = \partial_\mu B_\nu -
\partial_\nu B_\mu$ the corresponding field tensor. The equations
for the scalar field $\phi$ and the gauge field $a_\mu$ remain
unaffected, but the equation for $\psi$ is now:
\be
{\tilde D}_\mu{\tilde D}^\mu \psi
= - {1 \over 4\omega} f^2 e^{-2|\psi|} { \psi \over |\psi |}
\ee
and for $B_\mu$
\be
\partial_\nu G^{\nu\mu}=j_\psi^\mu=ig[\psi^\star
{\tilde D}^\mu\psi - \psi ({\tilde D}^\mu\psi)^\star].
\ee
Under cylindrical symmetry these equations may be solved using the
same ansatz for $\phi$ and $a_\mu$ as before, and the counterparts
for the dielectric:  $\psi = \xi(r)e^{im\theta}$, and
$B_\theta = B(r)$ (with all other components of $B_\mu$ set to zero).
The new equations are
\bea\label{geq1}
{1\over r}{d\over dr}{\left(r{d\xi\over dr}\right)}
-{\left({m\over r} - gB\right)}^2 \xi + {f^2 e^{-2 \xi}
\over 4\omega}&=&0,\\
{d\over dr}{\left( {1\over r}{d\over dr}
(rB)\right)} + 2g{\left( {m\over r}- g B\right)}\xi^2&=&0.
\eea
Given that $f^2$ is confined to the (local) string core, the
asymptotics for the new fields are similar. While we still have
$\chi = \sqrt{-\frac{m^2}{2\lambda}}$ and  $a={n/ (r e_0)}$,
for $r\rightarrow \infty$,
we find that $\xi$ may go to any constant (if $e_0$ is the
asymptotic value of the electric charge, $\xi\rightarrow 0$), while
\be
B = {m \over r g}.
\ee
Overall we find that the dielectric behaves like a local string
superposed on the usual Nielsen--Olesen vortex. Its magnetic flux,
associated with the gauge field $B_\mu$, is quantized, with a
quantum number $m$.  This
is particularly curious as there is nothing  topological in the
nature of the
dielectric string. Somehow it borrows these features from the topological
nature of the $\phi$-string sourcing it.
The $\psi$ field does not have a potential, only kinetic terms plus a
source at the string.  Thus the $\psi$ field can take on any covariantly
constant value far away from the string, which amounts to constant
$|\psi|$, and a phase equal to $m\theta$.

Notice that none of the points made in Section~\ref{cosmo}, regarding
the cosmological implications of local strings coupled to an
ungauged dielectric, apply to the strings considered in this section.
The variations in $e$ experienced in the surroundings of these strings
are confined to microphysical distances, and have no direct cosmological
implications.

It could also happen that $\phi$ and $\psi$ are coupled to the same
$U(1)$ gauge field. Then
\be \label{l4}
{\cal L}=-(D^\mu\phi)^\star D_\mu\phi-V(\phi)
-{f_{\mu\nu}f^{\mu\nu}\over 4 \epsilon^2}
 - \omega
(D_\mu\psi)^\star(D^\mu\psi),
\ee
leading to equations:
\bea
D_\mu D^\mu\phi&=&{\partial V\over \partial \phi^\star},\\
D_\mu D^\mu \psi
&=& - {1 \over 4\omega} f^2 e^{-2|\psi|} { \psi \over |\psi |},
\eea
and the gauge-field equation:
\bea
\partial_\nu {f^{\mu\nu}\over \epsilon^2} &=& j_\phi^\mu +j_\psi^\mu,
\\
j_\phi^\mu &=& ie_0[\phi^\star
D^\nu\phi - \phi (D^\nu\phi)^\star],\\
j_\psi^\mu &=& ie_0[\psi^\star
D^\nu\psi - \psi (D^\nu\psi)^\star].
\eea
Studying the asymptotics of these equations we find that
in this case the quantum number $m$ associated with $\psi$ would
have to be the same as $n$. Indeed we have that
$\phi=\chi(r)e^{in\theta}$,
with $\chi = \sqrt{-\frac{m^2}{2\lambda}}$, and $a={n/ (r e_0)}$;
but now we should also have $\psi = \xi(r)e^{in\theta}$ with
$\xi(r)$ going to any constant. More generally it could be that the
charge of the $\psi$ field is $g=ke$, where $k$ is an integer (or more
generally a rational number), in which case $m=kn$.

\section{Concluding remarks}\label{conc}
In this paper we studied the counterpart of the Nielsen--Olesen
vortex in Bekenstein's varying-$\alpha$ theory, by means of
analytical asymptotic methods, and numerically. We found that
such strings are covered by the dielectric medium characterizing
Bekenstein's theory. This coating, in effect, looks like a global
string superposed upon the local string core. The electric charge would
thus vary (typically increase) as the string core is approached.

We then discussed possible cosmological implications of such strings.
Clearly their networks will be much more complex
than just the superposition of a  local and a global
string network. We pointed out the main aspects in which
their dynamics and energy loss mechanisms will be more
complex. Structure formation in these theories will also have
more to it than just a superposition of results known to be true
for the two types of network. In addition we highlighted a
peculiar feature of these networks: their ability to generate
inhomogeneities in the electric charge, and consequently (among
other things) to generate inhomogeneous reionization scenarios.

In a brief section we compared these strings with fast-tracks:
solitonic solutions to VSL theories along which fast travel without
time-dilation effects is possible. We showed that while in some sense
fast travel along these strings is possible, in those cases
one cannot evade a
time-dilation effect. Curiously enough this time dilation
effect is present even
if observers do not exceed non-relativistic speeds. It is an effect
merely due to the fact that the pace of atomic clocks depends
upon $\alpha$, and slows down accordingly near the string core.

Finally we initiated an exploration of other solitonic solutions
in these theories. We considered the possibility that the dielectric
field itself might be a gauged. We found that in such a case,
even if the dielectric is not endowed with a potential, it
acquires topological features, e.g. quantization of
its associated magnetic field flux in the string core.

Although in this paper we restricted ourselves to gauged $U(1)$
symmetries it is possible to generalize our constructions to
non-abelian symmetry groups. Indeed counterparts to Bekenstein's
theory associated with strong interactions were discussed in
\cite{strong}. Following such generalizations it would be possible
to construct monopoles and textures (associated with $O(3)$
and $SU(2)$ gauge symmetries), covered by similar dielectric coatings.
The only type of soliton which apparently could not be associated
with changing-charge theories are domain walls, for which there is
no associated gauge symmetry. We defer to future work the scrutiny
of these more complicated solitons.

\section*{ACKNOWLEDGEMENTS}
We are grateful to J. Blanco--Pillado for very interesting comments,
and to Warren Perkins for supplying us with an integration
code, and for advice on numerics.
H.B. Sandvik would like to thank The Research Council of Norway for
financial support. This work was performed
on COSMOS, the Origin 2000 supercomputer owned by the UK-CCC and
supported by HEFCE and PPARC.

\end{document}